\documentclass{aa}
\usepackage{graphicx}
\usepackage{txfonts}
\usepackage{orcidlink}
\usepackage{threeparttable}
\usepackage{booktabs}
\usepackage{adjustbox}
\usepackage{amsmath} 
\usepackage{hyperref} 
\usepackage{float}
\usepackage{tabularx}
\usepackage{array}
\usepackage{CJKutf8}
\usepackage{multirow}
\usepackage{amssymb}
\usepackage{soul}
\usepackage{xcolor}
\usepackage[utf8]{inputenc}
\usepackage{hyperref}
\hypersetup{
    colorlinks=true,
    linkcolor=blue,
    citecolor=blue,
    filecolor=blue,
    urlcolor=blue
}
\begin{document} 
    \title{Group pre-processing in J1611+4026: Minor rejuvenation of a massive ETG fueled by interaction-driven gas transfer}
    \titlerunning{Group Pre-processing in J1611+4026}
    \author{
           Yaosong Yu  (\begin{CJK}{UTF8}{gbsn}{于耀淞}\end{CJK}) \inst{\ref{Westlake},\ref{YNU_SWIFAR}} \orcidlink{0009-0003-7681-3702} \thanks{E-mail: ayuyaosong@gmail.com}
           \and
           Qihang Chen (\begin{CJK}{UTF8}{gbsn}{陈启航}\end{CJK}) \inst{\ref{BNU_PA},\ref{BNU_FIAA}} \orcidlink{0009-0006-9345-9639}
           \and
           Zizhao He (\begin{CJK}{UTF8}{gbsn}{何紫朝}\end{CJK}) \inst{\ref{NCU_DOP},\ref{NCU_CRAHEP},\ref{PMO}} \orcidlink{0000-0001-8554-9163}
           \and
           Limeng Deng (\begin{CJK}{UTF8}{gbsn}{邓力艋}\end{CJK}) \inst{\ref{TUM},\ref{MPI}} \orcidlink{0009-0009-9255-920X}
            \and \\
           Liang Jing (\begin{CJK}{UTF8}{gbsn}{荆亮}\end{CJK}) \inst{\ref{BNU_PA},\ref{BNU_FIAA}}
           \orcidlink{0000-0003-1188-9573}
           \and
           Jianhui Lian (\begin{CJK}{UTF8}{gbsn}{连建辉}\end{CJK}) \inst{\ref{YNU_SWIFAR}}\orcidlink{0000-0001-5258-1466} \thanks{E-mail: jianhui.lian@ynu.edu.cn}}

\institute{
Department of Astronomy, Westlake University, Hangzhou 310024, People’s Republic of China \label{Westlake}
\and
South-Western Institute for Astronomy Research, Yunnan University, Kunming 650500, China \label{YNU_SWIFAR}
\and
School of Physics and Astronomy, Beijing Normal University, Beijing 100875, China \label{BNU_PA}
\and
Institute for Frontier in Astronomy and Astrophysics, Beijing Normal University, Beijing, 102206, China \label{BNU_FIAA}
\and
Department of Physics, Nanchang University, Nanchang, 330031, China \label{NCU_DOP}
\and
Center for Relativistic Astrophysics and High Energy Physics, Nanchang University, Nanchang, 330031, China \label{NCU_CRAHEP}
\and
Purple Mountain Observatory, Chinese Academy of Sciences, Nanjing, Jiangsu, 210023, China \label{PMO}
\and
Technical University of Munich, TUM School of Natural Sciences, Physics Department,  James-Franck-Stra{\ss}e 1, 85748 Garching, Germany
\label{TUM}
\and
Max Planck Institute for Astrophysics, Karl-Schwarzschild-Stra{\ss}e 1, 85748 Garching, Germany
\label{MPI}
}
\date{Received xxxx; accepted xxxx}

\abstract{
Satellite interactions can redistribute gas within galaxy groups, but their role in rejuvenating massive hosts remains difficult to establish observationally. J1611+4026 is a unique triple system at $z\simeq0.14$, comprising a massive early-type galaxy (component C) and two interacting star-forming satellites (components A and B). Combining DESI-LS imaging data with DESI and P200 spectroscopy, we perform image decomposition with \textsc{GALIGHT}, SED modeling with \textsc{CIGALE}, and stellar population analysis with \textsc{BAGPIPES}. An asymmetric tidal tail extending from A provides clear morphological evidence for the satellite interaction. Adopting the $g$-, $r$-, and $z$-band stellar mass-to-light ratio of A, we estimate a total stellar mass of $\sim1.41\times10^{9}\,M_{\odot}$ for the detected portion of the tail. Although dominated by an old stellar population, Component C exhibits a near-ultraviolet excess and a substantial dust attenuation of $A_V=1.05\pm0.04\,\mathrm{mag}$. Its star formation history, reconstructed from the optical spectrum, indicates renewed star formation over the past $\sim100\,\mathrm{Myr}$ following an extended period of suppressed activity. This episode contributes less than $0.1\%$ of the total stellar mass, characterizing it as minor rejuvenation. We infer that chemically enriched gas stripped during the A--B encounter subsequently fuels this activity in component C. As an external analogue of the Milky Way--Large Magellanic Cloud--Small Magellanic Cloud system, J1611+4026 allows us to examine together the satellite interaction and the recent stellar activity of the primary galaxy, and provide an observational benchmark for models of satellite-driven rejuvenation in massive galaxies.}


\maketitle

\section{Introduction}\label{Section1}
Galaxy interactions and mergers play an important role in the hierarchical assembly and evolution of galaxies. Tidal forces can produce tidal tails and reshape galactic morphologies \citep{Toomre1972, Duc2013}. In gas-rich encounters, gravitational torques can redistribute angular momentum and drive gas toward galactic centers \citep{Barnes1996}, potentially triggering starbursts and fueling active galactic nuclei (AGNs) \citep{Sanders1988, Hopkins2006}. Observational studies of galaxy pairs demonstrate enhanced star formation rates (SFRs) relative to control samples \citep{Ellison2008, Scudder2012}. Mean SFR enhancements can reach a factor of approximately three at small projected separations, while weaker enhancements remain detectable out to projected separations of approximately $150\,\mathrm{kpc}$ \citep{Patton2013}. Galaxy groups provide an environment where interactions among member galaxies can change their gas distributions and star formation activity. Environmental transformations occurring within groups before galaxies are accreted into more massive clusters are commonly referred to as pre-processing \citep{Fujita2004}. Studies of interacting dwarf galaxies show that encounters between low-mass companions can reshape their morphologies, redistribute gas, and enhance star formation. At the same time, proximity to a massive primary galaxy can further influence their gas content and subsequent evolution \citep{Besla2012, Stierwalt2015}.

Massive early-type galaxies (ETGs) are typically dominated by old stellar populations and exhibit little ongoing star formation, motivating their conventional description as ``red and dead'' \citep{Renzini2006}. Nevertheless, ultraviolet (UV)-optical studies have revealed evidence for recent star formation in a subset of ETGs \citep{Yi2005, Kaviraj2007}, including optically red and massive systems \citep{Werle2020, Salvador-Rusinol2020, Pandey2024}. Such activity may represent rejuvenation, in which star formation resumes after a period of quiescence \citep{Chauke2019}. External gas accretion provides a possible fuel source, as indicated by gas--stellar kinematic misalignments and low gas-phase metallicities in some ETGs \citep{Davis2011, Davis2019}. Gas-rich minor mergers have been proposed as a mechanism for supplying this material and triggering rejuvenation in individual systems \citep{Wang2025}. However, linking a specific interaction to the onset of renewed star formation requires joint constraints on the origin of the gas and the interaction history.

We focus on minor rejuvenation, in which renewed star formation contributes only a small fraction of the stellar mass. The star-forming main sequence (SFMS) relates SFR to stellar mass in star-forming galaxies (SFGs) \citep{Popesso2023}. Reconstructing a galaxy's star formation history (SFH) can reveal whether star formation resumed after quiescence \citep{Tanaka2024}. Detecting such weak episodes remains challenging because evolved stars can produce UV emission and SFH reconstruction is model-dependent \citep{Leja2019, Werle2020}. J1611+4026 comprises a massive early-type host galaxy (component C) and two satellites (A and B) whose interaction is traced by a tidal tail. In large-scale surveys, weak rejuvenation in systems of this kind may be overlooked if the massive host is not identified and analysed separately from its companions. J1611+4026 thus provides an opportunity to investigate whether gas stripped during satellite--satellite interactions could have fueled star formation in the host.

The Milky Way (MW) and the Large and Small Magellanic Clouds (LMC and SMC) provide a nearby example of a massive host accompanied by an interacting satellite pair, with orbital studies favoring a first-infall scenario for the Clouds \citep{Besla2007, Kallivayalil2013}. Models of the LMC--SMC interaction show that tidal stripping can contribute to the formation of the Magellanic Stream \citep{Besla2012}. If the stripped gas reaches the Galactic disk, it could provide fuel for future star formation \citep{Fox2014}. Although this nearby system permits detailed kinematic studies, our location within the Galactic disk complicates the reconstruction of its global structure and integrated properties because of extinction, distance uncertainties, and viewing geometry \citep{Rix2013, Bland-Hawthorn2016}. Extragalactic analogues provide an external view of such systems, allowing us to investigate whether gas stripped during satellite--satellite interactions can fuel minor rejuvenation in a quiescent primary massive galaxy.

In this work, we present a detailed multi-wavelength study of J1611+4026, a triple galaxy system at $z \simeq 0.14$ whose configuration of a massive host and an interacting satellite pair resembles that of the MW--LMC--SMC system. By combining morphological decomposition, spectral energy distribution (SED) modelling, and SFH reconstruction, we quantify the stellar mass contribution of a recent rejuvenation episode in the massive early-type host and investigate whether gas stripped during the satellite--satellite interaction could have fueled this activity. This paper is organised as follows. \hyperref[Section2]{Section 2} describes the spectroscopic follow-up observations, ancillary public data, and analysis methods. \hyperref[Section3]{Section 3} presents the main results, including a quantitative analysis of the tidal tail. \hyperref[Section4]{Section 4} discusses the robustness of our modelling and alternative rejuvenation mechanisms. We summarise our conclusions in \hyperref[Section5]{Section 5}. Throughout this work, we adopt a flat $\Lambda$ cold dark matter ($\Lambda$CDM) cosmology with $H_0 = 70\,\mathrm{km\,s^{-1}\,Mpc^{-1}}$, $\Omega_{\rm M} = 0.3$, and $\Omega_\Lambda = 0.7$.

\section{Data and methods}\label{Section2}
\subsection{Target selection}\label{Section2.1}
The J1611+4026 system was serendipitously noticed during the compilation of early version (unpublished) of the MGQPC catalogue \citep{Chen2026MGQPC-I, Chen2026MGQPC-III}. The MGQPC catalogue is a quasar pair candidate catalogue obtained by cross-matching version 8.0 of the Million Quasar catalogue \citep[MQC,][]{Flesch2023MQCv8} with \textit{Gaia} DR3 \citep{GaiaDR3}. Before the building of the MGQPC, the MQC version 7.9 was used to cross-match with \textit{Gaia} DR3 for pilot testing. The J1611+4026 system initially consists of J1611+4026A and B, where component A comes from the MQC version 7.9 as an AGN candidate, while component B comes from \textit{Gaia} DR3 as a candidate for a quasar pair member. This system seems like an AGN pair with both host galaxies relatively obvious, and might be mutually interacting with each other. Therefore, we took a spectroscopic follow-up for this system.

\subsection{Spectroscopy}\label{Section2.2}
Low-resolution long-slit spectroscopic follow-up was carried out for J1611+4026A and B using the Double Spectrograph (DBSP) equipped on the 200-inch Hale Telescope (P200) at Palomar Observatory, California, on the night of 5 September 2024 (P.I. Zizhao He). The dichroic D-55 was used to obtain the spectra at the blue and red arms. The exposure time for both sources and both arms was 900 seconds. The 300 lines/mm grating blazed at 3990{\AA} was chosen for the blue arm, while the 316 lines/mm grating blazed at 7150{\AA} for the red arm. This provided a dispersion of 2.108{\AA}/pixel and 1.535{\AA}/pixel for the blue and red arms \citep{Oke1982}. The seeing during the exposure of this system was approximately $1.7^{\prime\prime}$, and a $1.5^{\prime\prime}$-wide slit was chosen. During the exposure, the slit was rotated 78{\degr} from north to east to simultaneously obtain the spectra of components A and B. In the spectrum of the red arm, there is a bad CCD region at the wavelength of approximately 5800{\AA}--6250{\AA}. Thus, the spectra of both sources are unavailable in this range.

The DESI Legacy Imaging Surveys \citep[DESI-LS,][]{Dey2019} data of the J1611+4026 system indicate the existence of several other galaxies around components A and B. Thanks to the enormous spectral library of DESI DR1\footnote{\url{https://data.desi.lbl.gov/doc/}} \citep{DESI2025dr1}, and using the Spectra Analysis and Retrievable Catalogue Lab (SPARCL) from the Astro Data Lab\footnote{\url{https://datalab.noirlab.edu/index.php}}, we retrieved spectroscopic data of sources near component A within 1 arcmin and discovered component C, with a redshift of $z=0.1403$. Its redshift seems consistent with those of components A and B, preliminarily indicating a possible triple interaction.

Spectral fittings for all three components were performed using \textsc{QSOFITMORE} \citep{Fu2021}, an advanced Python module developed from \textsc{PyQSOFit} \citep{Guo2018} and optimised for analysing quasar and galaxy spectra. Widely used in recent studies, this module has been extensively demonstrated to be useful and reliable for quasar and galaxy spectral fitting \citep[e.g.,][]{Fu2022, Pang2025, Liang2025}.

\subsection{Photometry}\label{Section2.3}
We employed the Python-based package \textsc{GALIGHT}\footnote{\url{https://github.com/dartoon/galight}} \citep{Ding2020} to quantify the geometric parameters and surface brightness profiles of the three components, using the DESI-LS DR10 imaging data. The DESI-LS typically provides coverage in three optical bands ($g$, $r$, and $z$), and DR10 introduces additional $i$-band observations. However, we found that $i$-band data was unavailable for the sky region of the J1611+4026 system. Thus, we employed the $g$-, $r$-, and $z$-band data to composite the pseudo-colour image and perform photometric measurements and image decomposition.

Built upon the modelling engine \textsc{Lenstronomy} \citep{Birrer2021}, \textsc{GALIGHT} provides an automated framework for identifying source components, estimating background noise, and constructing Point Spread Functions (PSFs). It is capable of robustly disentangling complex light distributions, particularly in systems containing AGN or merger features. It has been extensively validated in various studies \citep[e.g.][]{Ding2023, Glazebrook2024, Jarvis2026}. Moreover, recent applications to complex interacting systems \citep[e.g.,][]{Wang2025} demonstrate its efficiency in recovering clean residual images. Given these attributes, \textsc{GALIGHT} was uniquely suited for quantifying the tidal features of J1611+4026A.

\subsection{SED modelling}\label{Section2.4}
We also used multi-band data to model the SEDs of the three components. These data include the GALEX NUV-band ultraviolet data \citep{Martin2005}, the Pan-STARRS $g$-, $r$-, $i$-, $z$-, and $Y$-bands optical data \citep{Chambers2016}, the 2MASS $J$-, $H$-, and $K_s$-bands near-infrared data \citep{Skrutskie2006}, and the WISE $W1$- and $W2$-bands data \citep{Wright2010}. For components A and B, due to the low spatial resolution of GALEX NUV-band and the lack of 2MASS photometry, only the Pan-STARRS ($g$, $r$, $i$, $z$, and $Y$) and the WISE ($W1$ and $W2$) data were used. For component C, all the bands aforementioned were used since component C is a single and near-infrared bright source. However, the WISE imaging of components A and B is slightly unresolved. Therefore, resolution-enhanced unWISE coadds data \citep{Lang2014, Schlafly2019} were used as the upper limits of $W1$ and $W2$ magnitudes of components A and B to ensure the robustness of the modelled SEDs. The Python-based SED fitting tool \textsc{CIGALE} \citep[Code Investigating GALaxy Emission,][]{Noll2009, Boquien2019} was employed to derive the physical properties of the three components.

\section{Results}\label{Section3}
\subsection{The J1611+4026 triplet}\label{Section3.1}
\begin{figure}
\centering
    \includegraphics[width=0.5\textwidth]{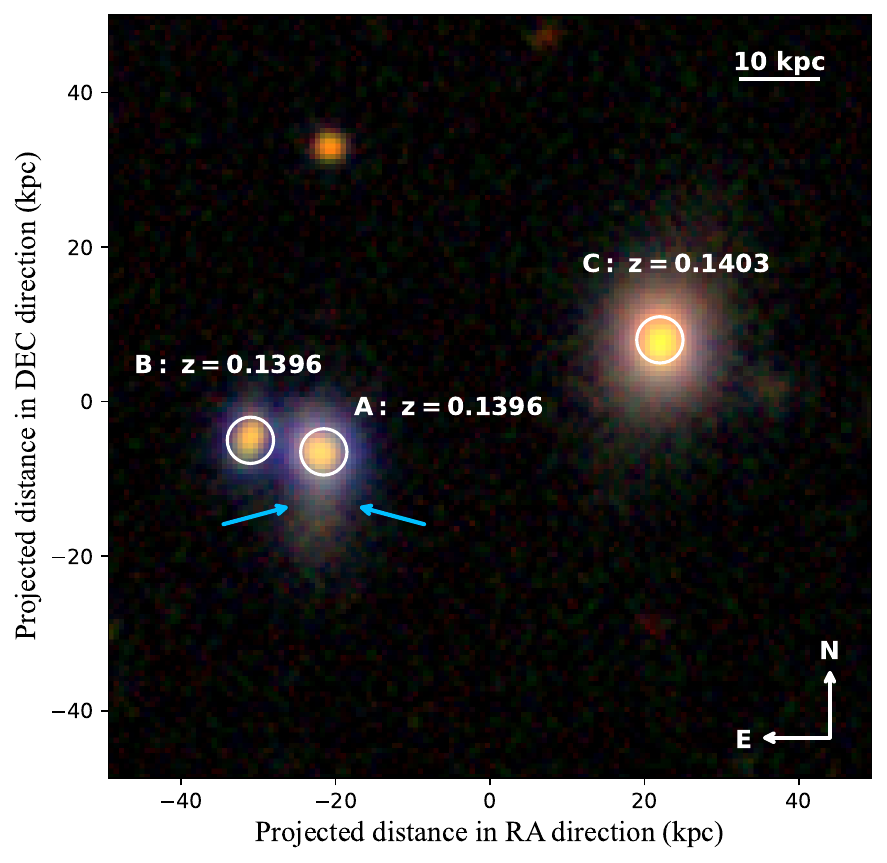} 
    \caption{The J1611+4026 system pseudo-colour image that composited from $g$-, $r$-, and $z$-band imaging data of DESI-LS DR10 using \texttt{HumVI} \citep{Marshall2015SWHumVI, Marshall2016HumVI} via RGB combination. The size of the image is $40\arcsec \times 40\arcsec$. The two blue arrows highlight the tidal tail of J1611+4026A. The white circles mark the three components, centred on the coordinates given by the DESI-LS DR10 catalogue.}
    \label{fig1}
\end{figure}

According to the DBSP/P200 spectroscopic follow-up of components A and B, and the DESI DR1 spectrum of component C, we preliminarily regarded the J1611+4026 system as a triple galaxy. \hyperref[fig1]{Figure 1} displays the DESI-LS DR10 pseudo-colour image of the J1611+4026 triple system, and highlights a faint tidal tail that extends from component A (two blue arrows), indicating potential gravitational interaction. The basic properties of these components are listed in \hyperref[table1]{Table 1}. We calculated a radial velocity difference of 184.09 $\mathrm{km\,s^{-1}}$ between components A and C using Equation 10 of \citet{Hogg1999QPdvr} or Equation 1 of \citet{Chen2026MGQPC-III}, suggesting the three components are at the same redshift \citep{Hennawi2006QPdvr, Hennawi2010QPdvr}. A transverse distance of less than 30 kpc between two galaxies is generally considered a strong hint of merging, placing the system well within the regime of strong gravitational interaction \citep{Ellison2010}. In addition, \citet{Patton2013} found that the mean star formation enhancement in their SDSS galaxy-pair sample extends to projected separations of $\sim150$~kpc, with no net enhancement detected at larger separations. Based on the J2000 coordinates and redshifts of the three components, we calculated a transverse distance of 9.038 kpc between components A and B, and that of 46.352 kpc between component A and C, suggesting that the entire J1611+4026 system is plausibly interactive. Consequently, we consider components A and B to be currently merging and will likely subsequently interact with component C. However, verifying these dynamical hypotheses requires precise 3-Dimensional kinematic measurements or intergalactic gas observations \citep[e.g.,][]{Karera2022, Glowacki2024}.

\begin{table}
    \centering
    \renewcommand{\arraystretch}{1.5}
    \setlength{\tabcolsep}{8pt}    
    \caption{Basic features of J1611+4026 system.}
    \label{table1}
    \begin{tabular}{lcccc}
        \hline \hline 
        Name & R.A. & Dec & $r$-band & redshift\\
         & (deg) & (deg) & (mag) & \\
        \hline
        J1611+4026A & 242.9772 & 40.4359 & 18.16 & 0.1396\\
        J1611+4026B & 242.9785 & 40.4362 & 19.58 & 0.1396\\
        J1611+4026C & 242.9707 & 40.4375 & 17.36 & 0.1403\\
        \hline
    \end{tabular}
    \tablefoot{The R.A., Dec, and $r$-band magnitudes are retrieved from the DESI-LS DR10 imaging catalogue. The redshifts of components A and B are measured by their DBSP/P200 spectra, while the redshift of component C is given by DESI DR1.}
\end{table}

\begin{figure*}
\centering
    \includegraphics[width=\textwidth]{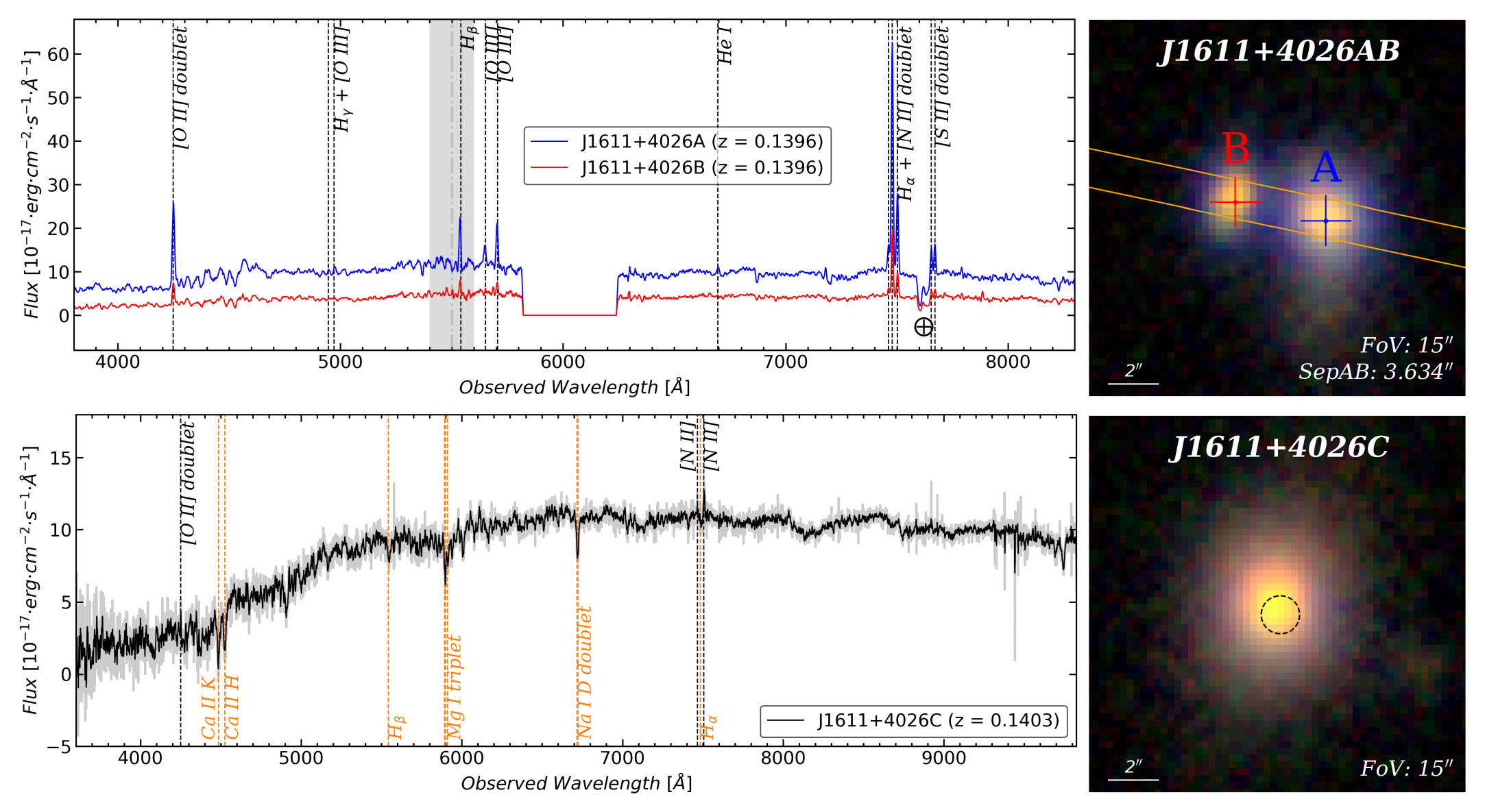}
    \caption{Upper panel: DBSP/P200 spectra of components A and B, with major emission lines (orange dashed lines) overlapping on it. The grey region around 5500{\AA} denotes the splicing of spectra from blue and red arms, while the black open circle with a cross denotes telluric absorption at $\sim$ 7600{\AA}. The $15\arcsec \times 15\arcsec$ pseudo-colour cutout image is shown on the right, with a rotated spectroscopic slit (two orange lines) overlapping on it (right). Lower panel: DESI DR1 spectrum of component C, with several weak emission lines (black dashed lines) and major absorption lines (orange dashed lines) overlapping on it. Its $15\arcsec \times 15\arcsec$ pseudo-colour image is shown on the right. The black dashed circle with a diameter of 1.5{\arcsec} denotes the fibre diameter and position during the observation of the DESI. Both pseudo-colour images, same as in \hyperref[fig1]{Figure 1}, are composited from $g$-, $r$-, and $z$-band images of DESI-LS DR10 using \texttt{HumVI}.}
    \label{fig2}
\end{figure*}

The optical spectra of the three components and their pseudo-colour cutout images were displayed in \hyperref[fig2]{Figure 2}. Components A and B are characterised by obvious emission features, notably strong H$\alpha$, H$\beta$, and prominent oxygen lines (e.g., [O~II] and [O~III]). Both continua exhibit typical spectral features of SFGs, despite their colours in the DESI-LS images not being the characteristic blue typically associated with SFGs and their lack of spiral arms in morphology. In contrast, the spectrum of component C differs markedly, exhibiting features of a passive galaxy, with only weak [N~II] and [O~II] emission. In addition, the spectrum is characterised by strong stellar absorptions (e.g., Ca~II K \& H, Mg~I triplet, and Na~I D doublet). The H$\alpha$ and H$\beta$ lines may indicate intrinsic weak absorption, while it is also possible that the intrinsic emission is weaker than the absorption, resulting in an overall weak absorption profile \citep[``lineless'' retired galaxies, see][for details]{Herpich2018}.

Spectral fittings for the three components were performed using \textsc{QSOFITMORE} to preliminarily derive their SFR. The fitting results were displayed in \autoref{AppendixA}. Both components A and B were fitted successfully, especially at their major emission lines. However, the fitting of component C was unreliable, as it may be a "lineless" retired galaxy, and the fitting performed by \textsc{QSOFITMORE} along with the selected templates might not be suitable for this component -- even though the fit appears satisfactory. Instead, we derived the SFR of component C from broadband SED fitting using \textsc{CIGALE} (see \hyperref[Section3.3]{Section 3.3}).

\subsection{Morphology of J1611+4026 system}\label{Section3.2}
\begin{figure*}
\centering
    \includegraphics[width=0.7\textwidth]{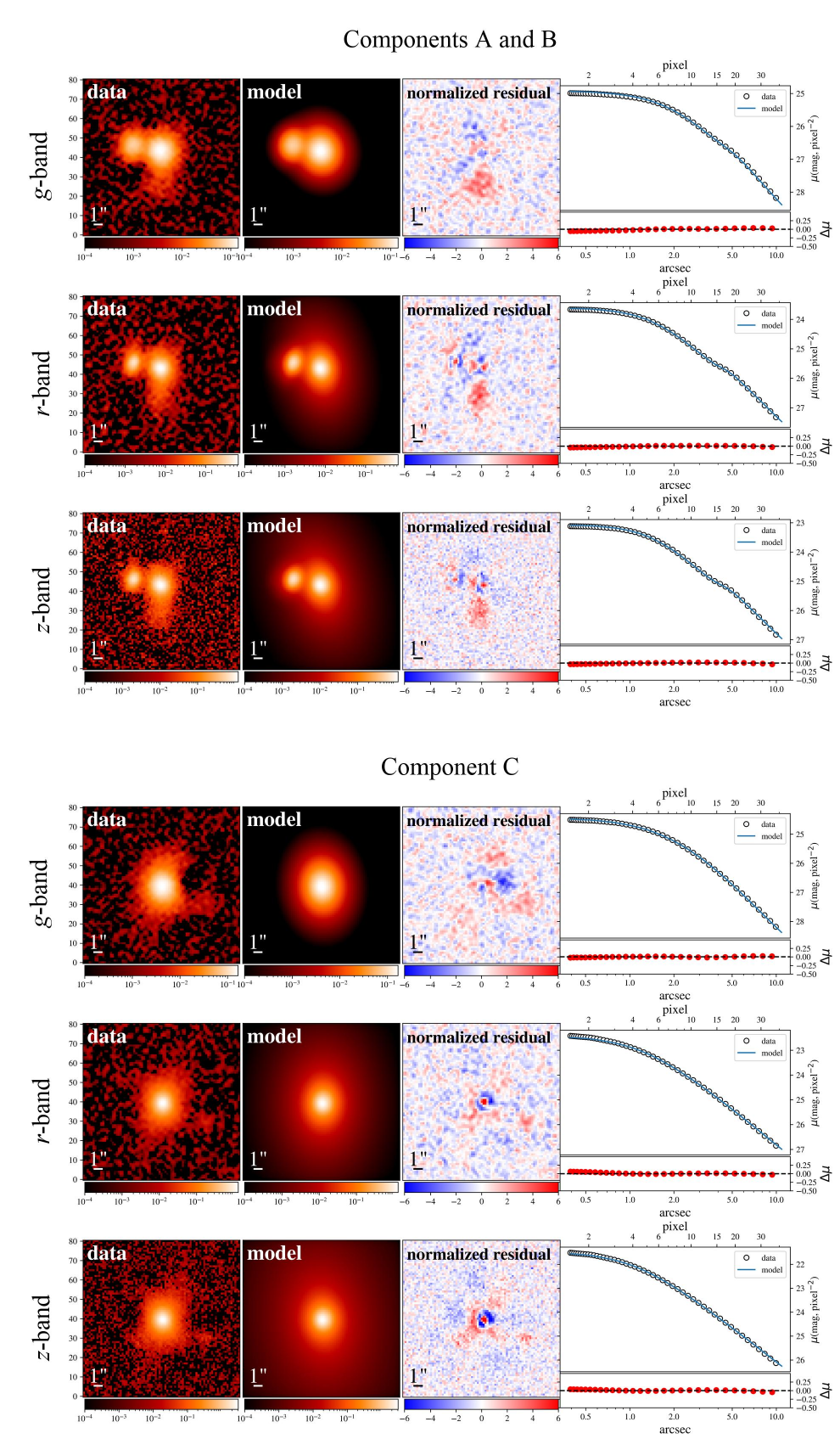} 
    \caption{Single-S\'ersic modeling of the three components in the DESI-LS DR10 $g$-, $r$-, and $z$-band images using \textsc{GALIGHT}. The left panels show the simultaneous fits to components A and B, and the right panels show the independent fits to component C. From top to bottom, the rows correspond to the $g$, $r$, and $z$-bands. Each set of panels shows the observed image, best-fit model, normalised residual map, and radial surface-brightness profiles with their differences. The localized central residuals in the $r$- and $z$-band fits of component C may reflect limitations of the adopted PSF models.}
    \label{fig3}
\end{figure*}

We modelled the DESI-LS DR10 $g$-, $r$-, and $z$-band images of the three components using \textsc{GALIGHT}. Owing to their small separation and slight blending, components A and B were fitted simultaneously. Each of them was fitted with a single S\'ersic profile. Component C was fitted separately with a single S\'ersic profile. For component B, allowing the S\'ersic index to vary in the $z$-band led to parameter degeneracies and unstable convergence. We therefore fixed $n = 1.00$ in this band, corresponding to an exponential profile. As shown in \hyperref[fig3]{Figure 3}, the models reproduce the overall light distributions and radial surface-brightness profiles. However, the normalised residual maps of component C show abnormal structures in the centre, especially in the $r$- and $z$-bands. These structures would be due to the mismatch of the point-spread function (PSF) model. \hyperref[tab2]{Table 2} summarises the best-fit parameters for the three components in all three bands. The reduced chi-square values, $\chi_r^2=0.99$--$1.25$, are close to unity, supporting an overall satisfactory fit under the adopted noise model. The fitted effective radii of components A and B decrease from the $g$-band to the $z$-band, indicateing a more compact light distribution at longer wavelengths. For component A, the accompanying increase in the S\'ersic index ($n = 0.91$--$4.57$) is consistent with a relatively compact, redder central component and more extended, bluer emission. Component B has a small fitted effective radii ($R_e = 0.16$--$0.63\,\mathrm{kpc}$) and a low S\'ersic indices ($n = 0.44$--$1.76$), characterizing it as a compact source with a low-index light profile. Component C shows a decrease in effective radius from $R_e = 1.05$ to $0.73\,\mathrm{kpc}$ and an increase in S\'ersic index from $n = 1.08$ to $6.28$ across the $g$- and $z$-bands. These trends suggest a more compact light distribution at longer wavelengths, and a redder central component.

\begin{table*}
\label{tab2}
    \centering
    \renewcommand{\arraystretch}{1.4} 
    \setlength{\tabcolsep}{12pt}        
    \begin{threeparttable}
        \caption{\textsc{GALIGHT} best-fit morphological properties of the three components.}        
        \begin{tabular}{lcccc}
            \hline \hline
            Parameter & Symbol & component A & component B & component C \\
            \hline
            \multicolumn{5}{c}{\textit{g}-band} \\
            \hline
            Effective radius & $R_{e,g}$ (kpc) & $0.87 \pm 0.01$ & $0.63 \pm 0.01$ & $1.05 \pm 0.01$ \\
            S\'ersic index & $n_{\text{s\'ersic},g}$ & $0.91 \pm 0.02$ & $0.44 \pm 0.03$ & $1.08 \pm 0.02$ \\
            Reduced $\chi^2$ & $\chi^2_r$ & 0.99 & 0.99 & 1.02 \\
            \hline            
            \multicolumn{5}{c}{\textit{r}-band} \\
            \hline
            Effective radius & $R_{e,r}$ (kpc) & $0.60 \pm 0.01$ & $0.21 \pm 0.03$ & $0.97 \pm 0.01$ \\
            S\'ersic index & $n_{\text{s\'ersic},r}$ & $3.21\pm0.10$ & $1.76 \pm 0.35$ & $3.63 \pm 0.06$ \\
            Reduced $\chi^2$ & $\chi^2_r$ & 1.23 & 1.23 & 1.01 \\
            \hline            
            \multicolumn{5}{c}{\textit{z}-band} \\
            \hline
            Effective radius & $R_{e,z}$ (kpc) & $0.47 \pm 0.01$ & $0.16 \pm 0.02$ & $0.73 \pm 0.01$ \\
            S\'ersic index & $n_{\text{s\'ersic},z}$ & $4.57\pm0.24$ & $1.00\pm0.00^{\text{*}}$ & $6.28 \pm 0.16$ \\
            Reduced $\chi^2$ & $\chi^2_r$ & 0.99 & 0.99 & 1.25 \\
            \hline
        \end{tabular}
        
        \begin{tablenotes}
            \footnotesize
            \item \textbf{Note.}
            \item[$^{\text{*}}$] Fixed to 1.0 to ensure fitting robustness and stable convergence. (see \hyperref[Section3.2]{Section 3.2}).
        \end{tablenotes}
    \end{threeparttable}
\end{table*}

\subsection{Tidal tail mass estimation}\label{Section3.3}
\begin{figure*}
\centering
    \includegraphics[width=1\textwidth]{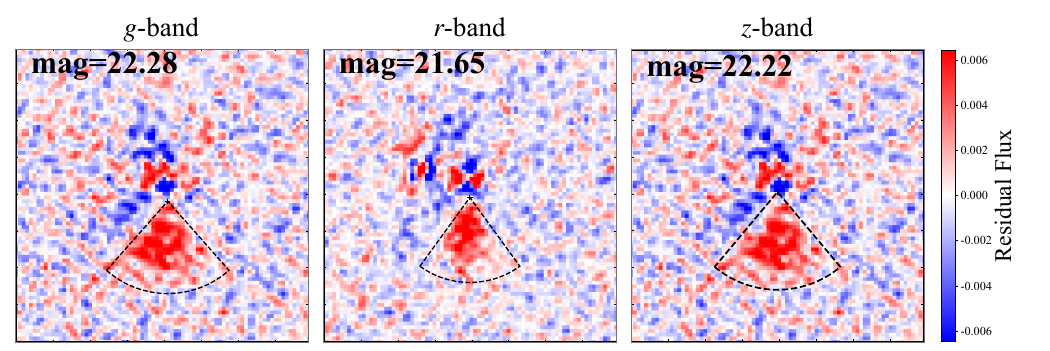} 
    \caption{Unnormalised $g$-, $r$-, and $z$-band residual maps (in units of $\sigma$) and the fan-shaped apertures (black dashed lines). The tidal tail was measured to have an integrated magnitude of $m_g = 22.28$, $m_r = 21.65$, and $m_z = 22.22$. All the maps share the same colorbar.}
    \label{fig4}
\end{figure*}

The residual maps of components A and B in \hyperref[fig3]{Figure 3} reveal a faint, asymmetric feature extending from component A, which we interpret as a tidal tail. We measured its $g$-, $r$-, and $z$-band fluxes using fan-shaped apertures set by visual inspection. Note that the areas of the three apertures are slightly different (not completely equal) according to the actual residual distribution. \hyperref[fig4]{Figure 4} shows the unnormalised residual maps of components A and B, with corresponding apertures overlapping on it. Photometry was performed on the unnormalised data-minus-model residuals. The resulting magnitudes of the tail are $g = 22.28$, $r = 21.65$, and $z = 22.22$.

The tail stellar mass was estimated by \textsc{CIGALE} adopting the band-dependent stellar mass-to-light ratios of component A. 
According to the SED fitting, the total mass of component A was $(1.46\pm0.25)\times10^{10}\,M_\odot$. Assuming that the tail lies at the same distance as component A and shares the same mass-to-light ratio in each band, we scaled the stellar mass of component A by the corresponding tail-to-galaxy flux ratio. Given the DESI-LS DR10 photometry of component A ($g_A = 18.8416$, $r_A = 18.1570$, and $z_A = 17.6210$), the estimated tail stellar mass are $6.15\times10^{8}\,M_\odot$, $5.85\times10^{8}\,M_\odot$, and $2.11\times10^{8}\,M_\odot$ in the $g$-, $r$-, and $z$-band, respectively. Note that these mass estimations excluded diffuse and faint emission outside the apertures. Besides, emissions at other bands are not take into consideration. Therefore, the total tail mass is still underestimated.

\subsection{Physical properties of J1611+4026 system}\label{Section3.4}
\begin{figure*}
\centering
    \includegraphics[width=1\textwidth]{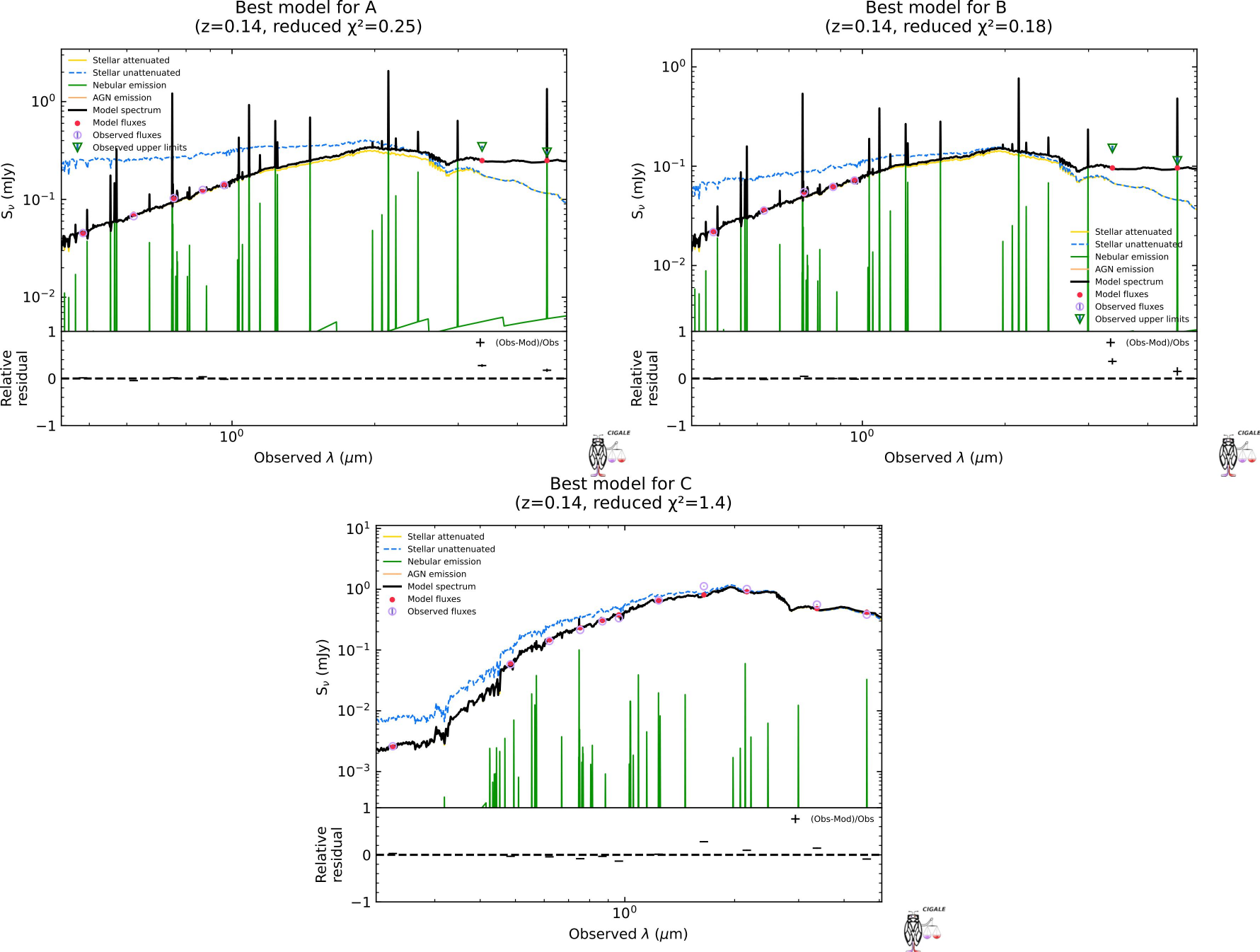} 
    \caption{Best-fit SEDs for components A, B, and C modelled by \textsc{CIGALE}. Green downward-pointing triangles denote WISE upper limits for A and B, whose individual infrared emission cannot be reliably separated at the WISE resolution. The lower panels are relative residuals.}
    \label{fig5}
\end{figure*}

The input parameters of SED modelling by \textsc{CIGALE} are listed in \hyperref[tab_SED]{Table C1}, while the best-fitting models are shown in \hyperref[fig5]{Figure 5}. Components A and B cannot be reliably separated under the angular resolution of WISE, even though we applied unWISE. We therefore treated their WISE photometry as upper limits on the infrared emission from each component (green downward triangles). The models reproduce the photometric data well and are consistent with the upper limits of WISE photometry aforementioned. The reduced chi-square values are $\chi_r^2 = 0.25$, $0.18$, and $1.4$ for components A, B, and C, respectively. \hyperref[tab3]{Table 3} summarises the physical properties inferred from the spectroscopy and SED modelling. The SED-derived stellar masses are $(1.46\pm0.25)\times10^{10}$, $(0.75\pm0.16)\times10^{10}$, and $(5.37\pm1.40)\times10^{10}\,M_\odot$ for A, B, and C, respectively, indicative of component C as the most massive one. 

\begin{table*}
\centering
\renewcommand{\arraystretch}{1.5}
\setlength{\tabcolsep}{4pt}
\setcounter{table}{2}
\begin{threeparttable}
\caption{Primary results of spectral fitting and SED modelling for three components.}
\label{tab3}
    \begin{tabular}{c|ccc|ccc}
        \hline \hline
         & \multicolumn{3}{c|}{\textsc{QSOFITMORE}} & \multicolumn{3}{c}{\textsc{CIGALE}}\\
        \hline
        Component & H$\alpha$ flux & $\mathrm{SFR}_{\mathrm{H}\alpha}$
    & Metallicity (N2)\tnote{*} & $\mathrm{SFR}_{\mathrm{SED}}$ & $M_{\star, \mathrm{SED}}$ ($\times\, 10^{10} M_{\odot}$) & $E(B-V)_{\mathrm{SED}}$ \\
        \hline
        A & $499.6 \pm 10.60$ & $6.37 \pm 1.10$ \tnote{\textdagger} & $8.66 \pm 0.006$ & $4.22$ & $1.46 \pm 0.25$ & $0.54 \pm 0.16$ \\
        B & $134.7 \pm 3.31$ & $>0.591 \pm 0.02$ & $8.67 \pm 0.012$ & $1.85$ & $0.75 \pm 0.16$ & $0.45 \pm 0.18$ \\
        C & - & - & - & 0.58 & $5.37 \pm 1.40$ & $0.53 \pm 0.15$ \\
        \hline
    \end{tabular}
    \begin{tablenotes}
        \footnotesize
        \item[*] Derived by empirical `N2' method \citep{Pettini2004}.
        \item[\textdagger] Derived by attenuation corrected H$\alpha$ flux.
        \item \textbf{Note.} For component C, the low spectral signal-to-noise ratio precluded a robust fit to the H$\alpha$ emission. We were unable to derive reliable H$\alpha$ flux and metallicity.
    \end{tablenotes}
\end{threeparttable}
\end{table*}

To exclude the possibility of AGN contribution, we included the \textsc{SKIRTOR} model \citet{Stalevski2016} in \textsc{CIGALE} and allowed the AGN fraction to vary from 0.0 to 0.99. The AGN fraction is defined as the AGN contribution to the modelled infrared luminosity. The inferred AGN fractions are $f_{\rm AGN}=0.075$, $0.042$, and $0.018$ for A, B, and C, respectively. These estimation favor a small AGN contribution within the \textsc{SKIRTOR} model.

We estimated the SFR of the three components according to the fitting of the spectra (\textsc{QSOFITMORE}) or SED (\textsc{CIGALE}). Assuming an intrinsic H$\alpha$/H$\beta$ ratio of 2.86 \citep{Groves2012} and applying the attenuation law proposed by \citet{Calzetti2000}, we corrected the attenuated H$\alpha$ flux for component A using the Balmer decrement. Furthermore, applying the calibration proposed by \citet{Kennicutt1998} for a Salpeter initial mass function (IMF) gives an SFR of $6.37\,M_\odot\,\mathrm{yr}^{-1}$. For component B, a low signal-to-noise ratio measurement of H$\beta$ yields a Balmer decrement below the assumed intrinsic ratio, preventing a reliable attenuation correction. We give an SFR of $0.591\,M_\odot\,\mathrm{yr}^{-1}$ (based on the uncorrected H$\alpha$ flux) as a lower limit for component B, assuming that all the H$\alpha$ emission originates from star formation. The SED fitting provides an SFR estimation of $1.85\,M_\odot\,\mathrm{yr}^{-1}$, which is reported here for reference. For component C, the H$\alpha$ emission-line measurement is unreliable. Instead, we give the SED-derived SFR of $0.58\,M_\odot\,\mathrm{yr}^{-1}$.

Both components A and B exhibit significant ongoing star formation, while the massive component C shows comparatively weak activity. For component C, the SED-derived SFR of $0.58\,M_\odot\,\mathrm{yr}^{-1}$ and stellar mass of $(5.37\pm1.40)\times10^{10}\,M_\odot$ imply a specific SFR of approximately $1.1\times10^{-11}\,\mathrm{yr}^{-1}$. This is consistent with a scenario in which weak recent star formation occurred in an originally quiescent massive galaxy.

\section{Minor rejuvenation}\label{Section4}
\subsection{Anomalous dust attenuation}\label{Section4.1}

Although many dust-bearing ETGs have been identified in the local Universe \citep{Rowlands2012}, the dust attenuation of component C, $E(B-V) = 0.53\pm0.15$, is noteworthy given the weak star formation hints in its spectrum. According to panel (a) of Figure \ref{fig6}, component C lies below the SFMS proposed by \citet{Popesso2023}. However, the position of a galaxy relative to the SFMS cannot uniquely determine its long-term stellar activity or SFH, as simulations show that galaxies may undergo repeated cycles of quenching and rejuvenation \citep{Fortune2025}.

The gas-phase oxygen abundances of components A and B are derived using the empirical N2 method \citep{Pettini2004}, where the N2 is defined as $\mathrm{N2}\equiv\log_{10}([\mathrm{N\,II}]\lambda6584/\mathrm{H}\alpha)$. Panel (b) of Figure \ref{fig6} indicates that both components have similar abundances and satisfy the local mass--metallicity relation proposed by \citep{Lian2018}, indicating chemically enriched gas in both components. In addition, in the rest-frame $UVJ$ color-color diagram \citep[panel (c) of Figure \ref{fig6},][]{Williams2009, Schreiber2015}, all three components are located in the star-forming region, with component C positioned further away from the quiescent region and exhibiting a significantly redder $V-J$ color than the other two. This would be due to potentially abnormal dust extinction. Given these findings, reconstructing its SFH is essential to distinguish persistent low-level star formation from renewed activity following a quiescent phase.

\subsection{Star formation history of component C}\label{Section4.2}
\begin{figure*}
\centering
    \includegraphics[width=\textwidth]{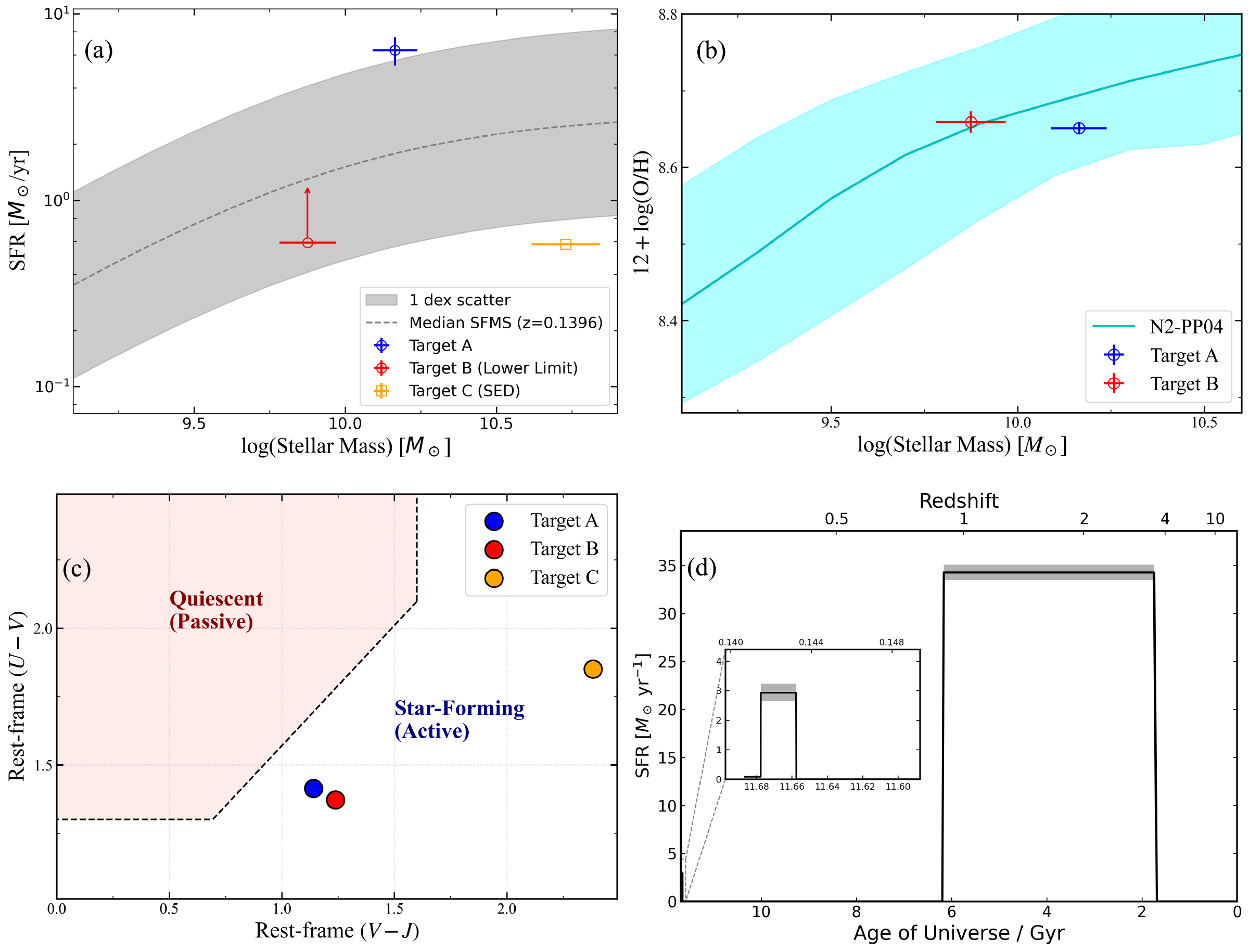} 
    \caption{Star formation diagnostics and reconstructed SFH of J1611+4026 triplet. Components A, B, and C are represented in blue, red, and orange symbols, respectively. (a) SFR versus stellar mass. The dashed line shows the SFMS relationship at $z \sim 0.14$ from \citet{Popesso2023}, and the gray band spans $\pm0.5$ dex around this relationship. The red unfilled circle with an upward red arrow represents the SFR lower limit of B. Note that we only present the SFRs of A and B inferred from the spectral fitting of H$\alpha$. The SED-derived SFRs are provided in the text solely for reference. The orange square represents the SED-derived SFR of C. (b) Gas-phase oxygen abundance versus stellar mass for A and B. The cyan curve and shaded area show the local reference relationship and its scatter, which is obtained by the N2--PP04 calibration \citep{Lian2018}. (c) Rest-frame $UVJ$ color-color diagram. The color indices of the three components are derived from the best-fitting broadband SEDs (see \hyperref[Section2.4]{Section 2.4}). The dashed lines are defined by \citet{Williams2009}, which separate the quiescent and star-forming galaxy selection regions at $0 < z < 0.5$. (d) SFH of component C reconstructed from the DESI optical spectrum using \textsc{BAGPIPES}. The horizontal axis gives the age of the Universe, with the corresponding redshift indicated above. The zoomed-in panel shows the recent SFH.}
    \label{fig6}
\end{figure*}

To investigate whether component C experienced renewed star formation, the 11-band photometry and DESI optical spectrum were fitted using \textsc{BAGPIPES} \citep{Carnall2018}, respectively. We adopted a nonparametric SFH model with eight time bins and a continuity prior. The input model parameters and priors are summarised in \hyperref[tab4]{Table 4}. The optical spectrum provides constraints on stellar ages and metallicity through the 4000~\AA\ break, Balmer absorption lines, and metal absorption features \citep{Conroy2013}. Both fitting approaches yielded similar results, so we display the SFH derived from the spectral fitting in panel (d) of Figure \ref{fig6}. The reconstructed SFH indicates that most of the stellar mass formed in an early phase of star formation, followed by an extended phase of low-level activity. A recent increase in star formation over the past $\sim 100\,\mathrm{Myr}$ is highlighted in the zoomed-in panel. The inferred SFR in the recent time bin is approximately $3\,M_\odot\,\mathrm{yr}^{-1}$ according to this panel. Note that this SFR represents the average in the corresponding bin, and should be distinguished from an instantaneous SFR.

\begin{table*}
    \centering
    \caption{Summary of the \textsc{BAGPIPES} input parameters}  
    \label{tab4}
    \renewcommand{\arraystretch}{1.3} 
    \begin{tabular}{llc}
        \hline \hline
        Component & Parameter & Configuration / Prior \\
        \hline
        Star Formation History & Model Type & Continuity (Non-parametric) \\
        & $\log(M_*/M_{\odot})$ & Uniform: $[6.0, 13.0]$ \\
        & Metallicity ($Z$) & Log-10: $[0.01, 2.5] Z_\odot$ \\
        & Time Bins & 8 bins ($0$ Myr -- $10$ Gyr) \\
        Nebular Emission & $\log U$ & $-3.0$ \\
        Dust Attenuation & Law & Calzetti \citep{Calzetti2000} \\
        & $A_V$ & Uniform: $[0.0, 4.0]$ \\
        Redshift & $z$ & 0.1403 \\
        \hline
    \end{tabular}
\end{table*}

Under the adopted SFH model, the recent star formation activity contributes only a small fraction of the total stellar mass ($f_{\star} < 0.1\,\%$). The renewed activity and its small stellar mass contribution suggest the occurrence of a minor rejuvenation. In this scenario, component C recently experienced quite limited star formation while its stellar mass remained dominated by the older population.

\section{Discussion}\label{Section5}
\subsection{Robustness of the rejuvenation signal}\label{Section5.1}
To assess the minor rejuvenation in component C, we examined the posterior distributions from the \textsc{BAGPIPES} fit to its DESI optical spectrum. We focused on two aspects: the consistency of the attenuation estimates compared with \textsc{CIGALE} and the dependence of the recent SFH on dust attenuation and stellar mass. The marginalized attenuation posterior yields $A_V=1.05^{+0.04}_{-0.04}$ mag, with uncertainties corresponding to the 16th and 84th percentiles.

We first compared this result with the attenuation from the broadband SED fit using \textsc{CIGALE}. The latter yields a nebular color excess of $E(B-V)_{\mathrm{lines}}=0.53\pm0.15$. Adopting $E(B-V)_{\mathrm{stars}}=0.44E(B-V)_{\mathrm{lines}}$ and the Calzetti attenuation curve with $R_V=4.05$ \citep{Calzetti2000}, we obtain a stellar continuum attenuation of $A_V=0.94\pm0.27$ mag. The uncertainty is propagated from the color excess with the conversion factors held fixed. This value agrees with the baseline attenuation inferred by \textsc{BAGPIPES}, indicating that the optical spectrum and 11-band photometry consistently favor attenuation of approximately $1$ mag under the prescription.

To quantify the recent change in star formation, we defined $\Delta_{\mathrm{recent}}$ as the logarithmic ratio of the mean SFR over the past $100\,\mathrm{Myr}$ to that in the preceding $100$--$300\,\mathrm{Myr}$ interval. The posterior yields $\Delta_{\mathrm{recent}}=9.05^{+4.38}_{-4.20}$ dex and strongly favors positive values. This large contrast reflects the very low SFR inferred for the preceding interval, while its broad posterior indicates that the enhancement factor is poorly constrained. The result supports renewed activity following a period of suppressed star formation. A large relative increase is compatible with the small recent stellar mass contribution inferred in \hyperref[Section4.2]{Section 4.2}, because the preceding activity provides a nearly negligible reference level.

The joint posterior distributions in \hyperref[AppendixB]{Appendix B} show weak monotonic correlations between $\Delta_{\mathrm{recent}}$ and both $A_V$ and the current stellar mass, with Spearman coefficients of $\rho_{\mathrm{s}}=0.053$ and $0.097$, respectively. The recent-to-prior SFR contrast therefore does not exhibit a correlation with either attenuation or stellar mass within the adopted model.

The reconstructed recent activity also shows temporal evolution within the youngest bins. The logarithmic SFR ratio between the $0$--$10$ and $10$--$30\,\mathrm{Myr}$ intervals is approximately $-1.55$, implying that the mean SFR in the youngest bin is about $2.8\%$ of that in the preceding bin. Thus, the fitted SFH favors a recent episode that had already weakened by the time of observation. Together with the small stellar mass contribution of the recent episode, these results support the interpretation of minor rejuvenation in component C.

\subsection{Alternative scenarios}\label{Section5.2}
To assess alternative explanations for the attribution to recent star formation in component C, we examined possible contributions from AGN emission and hot evolved stars associated with the ``UV upturn''. We then discuss the possible origin of the gas supply that could fuel the inferred rejuvenation.

\subsubsection{AGN activity}\label{Section5.2.1}
We first determine the WISE photometry for evidence of an AGN contribution. Emission from hot circumnuclear dust can produce a red mid-infrared continuum, motivating the $W1-W2\geq0.8$ AGN selection criterion in the Vega system \citep{Stern2012}. Component C has $W1-W2\ll0.8$ and therefore falls well below this color threshold. Its color is consistent with galaxy light dominating the W1 and W2 bands and does not indicate a prominent AGN hot-dust component.

The optical spectrum provides a complementary diagnostic. No broad components are detected in H$\alpha$ or H$\beta$, providing no evidence for an unobscured broad-line nucleus. The narrow emission lines are weak, which does not permit a secure classification of the ionizing source from diagnostic line ratios. Their weakness alone cannot distinguish photoionization by young stars from other mechanisms. In particular, LINER-like emission in red galaxies can be spatially extended and associated with ionizing sources distributed throughout the evolved stellar population \citep{Yan2012}.

An obscured nucleus could lack detectable broad optical lines while contributing infrared emission through the reprocessing of accretion radiation by circumnuclear dust \citep{Assef2013}. The blue W1--W2 color of component C offers no evidence for such a component dominating these bands. Consistently, the SED decomposition yields a low model-inferred AGN contribution to the infrared luminosity, $f_{\mathrm{AGN}}\simeq0.018$. WISE color selection is nevertheless less sensitive to nuclei whose emission is diluted by the host galaxy \citep{Assef2013}. Taken together, the infrared photometry, SED decomposition, and absence of broad optical lines disfavor a prominent AGN contribution to the observed broadband emission. A weak nucleus may still be present, but the available observations provide no positive evidence that AGN emission accounts for the ultraviolet excess.

\subsubsection{UV Upturn}\label{Section5.2.2}
Massive quiescent ETGs can exhibit a ``UV upturn,'' characterized by a rise in their ultraviolet spectra at wavelengths shorter than approximately $2500\,\mathrm{\AA}$. This emission is generally attributed to hot, evolved low-mass stars, particularly extreme horizontal-branch stars and their descendants, rather than to recent star formation \citep{Yi1998O, Connell1999}. We therefore consider whether such an evolved stellar population could contribute to the ultraviolet emission of component C.

\begin{figure}
\centering
    \includegraphics[width=0.5\textwidth]{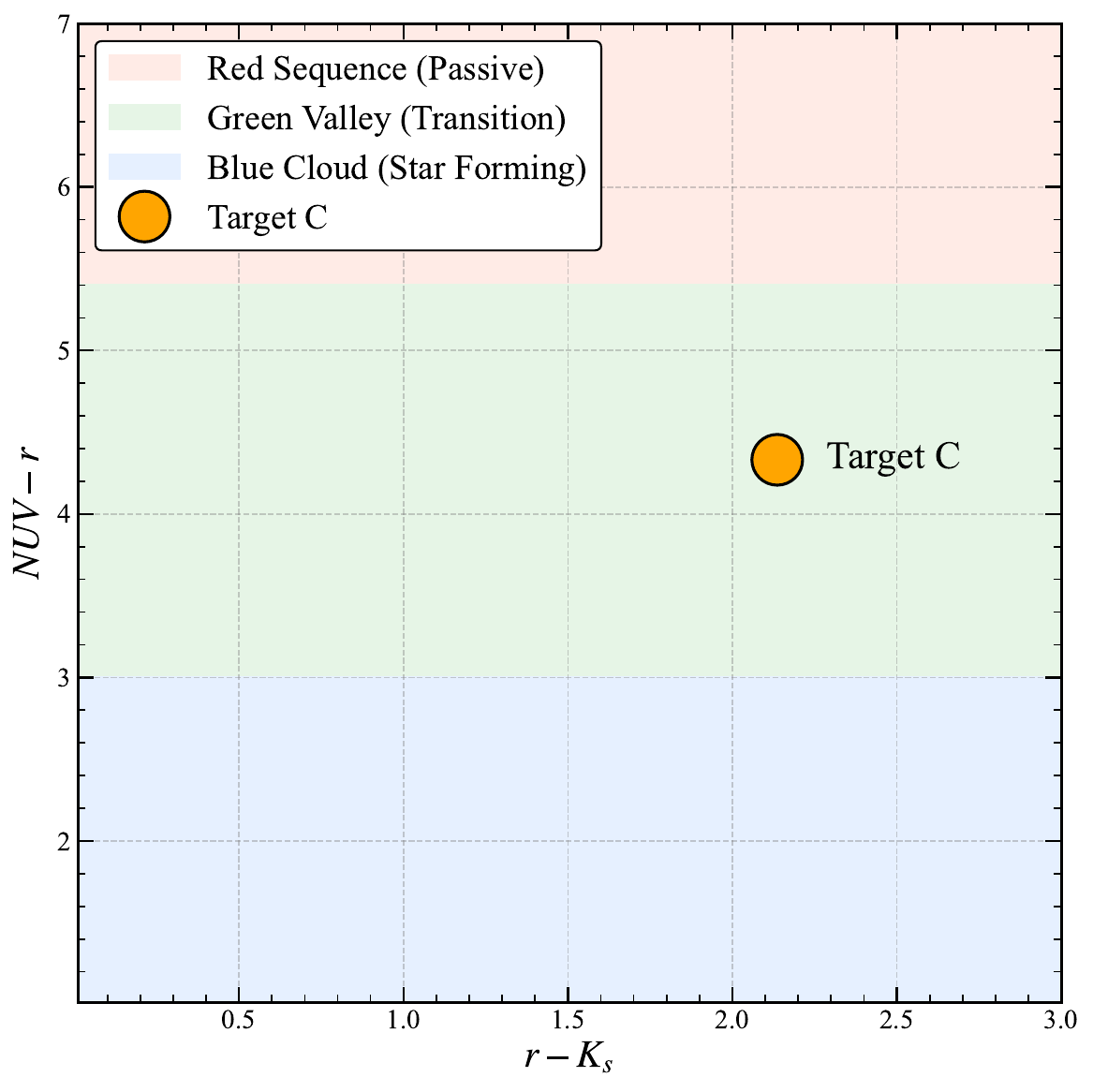} 
    \caption{Observed-frame $NUV-r$ versus $r-K_s$ colors of component C in the AB system. The orange circle marks the colors calculated from the SED-derived flux densities. Error bars show formal uncertainties propagated assuming independent errors between bands and are smaller than the marker. The blue, green, and pink shaded regions indicate the illustrative intervals $NUV-r<3$, $3<NUV-r<5.4$, and $NUV-r>5.4$, respectively. The boundary at $NUV-r=5.4$ corresponds to the empirical reference adopted by \citet{Schawinski2007} for nearby ETGs.}
    \label{fig7}
\end{figure}

\hyperref[fig7]{Figure 7} shows the observed-frame AB colors of component C, calculated from the SED-derived flux densities. We obtain $NUV-r=4.331\pm0.021$ and $r-K_s=2.136\pm0.021$, where the quoted formal uncertainties are propagated from the individual flux-density errors. The horizontal shaded regions indicate illustrative intervals in $NUV-r$. For comparison, \citet{Schawinski2007} adopted $NUV-r=5.4$ as an empirical boundary for identifying recent star formation in nearby ETGs. This reference is motivated by the synthetic color of NGC~4552, a classical elliptical galaxy with a strong UV upturn \citep{Kaviraj2007}. Component C is bluer than this reference, consistent with a contribution from recently formed stars. 

The near-ultraviolet color is consistent with the young stellar component inferred from the spectroscopic SFH reconstruction in \hyperref[Section4.2]{Section 4.2}. The color comparison therefore complements the SFH evidence for recent star formation while allowing an additional ultraviolet contribution from hot evolved stars.

\subsubsection{Tidal gas transfer and group Pre-processing}\label{Section5.2.3}
The interacting component pair A and B provides a plausible external source of the gas associated with the recent star formation in component C. The tidal tail extending from component A demonstrates that the encounter has redistributed material beyond its main stellar body, as expected during tidal interactions \citep{Toomre1972}. Both satellites also contain chemically enriched gas, as indicated by their oxygen abundances, as shown in panels (a) and (b) of \hyperref[fig6]{Figure 6}. Together with the dust attenuation and recent stellar activity inferred for component C, these properties motivate a scenario in which material stripped during the satellite--satellite encounter subsequently fuels star formation in the massive galaxy.

In this scenario, the interaction between A and B draws gas and dust into the surrounding environment, where part of the stripped material becomes available for accretion by C. Tidal debris can contain atomic gas, molecular gas, and dust in addition to stars \citep{Duc2013}. Gas captured by the massive galaxy can dissipate energy and redistribute angular momentum, allowing it to settle into regions where star formation can resume. Accretion of enriched material from the satellites would also supply dust, providing a physical explanation for the attenuation and recent star formation in component C. The proposed sequence therefore connects the dynamical evolution of the satellite pair to the renewed activity of the host before the satellites merge with it.

The relatively smooth optical morphology of C is compatible with this interpretation. Gas-rich tidal structures need not possess equally prominent stellar counterparts: observations of interacting groups reveal extended H~I bridges and streams that are much less conspicuous in optical images \citep{Duc2013}. Accretion dominated by gaseous material can therefore supply fuel without producing a comparably strong stellar tidal feature around the recipient. The small mass contribution of the recent episode is also consistent with a ``frosting'' of young stars superposed on an older population \citep{Trager2000}. Such an episode can modify the ultraviolet emission while contributing little to the total stellar mass, allowing C to retain its overall early-type appearance.

The unique feature of J1611+4026 is the possible connection between a two-satellite interaction and a weak stellar response in their primary massive galaxy. In the proposed picture, redistribution of gas within the group modifies the recent SFH of component C before substantial stellar accretion from the companions occurs. This offers a possible fueling pathway during group pre-processing, through which galaxies can evolve within smaller associations before their subsequent incorporation into more massive clusters \citep{Fujita2004}.

This interpretation makes specific predictions for the gas distribution and kinematics. Spatially resolved H~I observations could reveal a gaseous connection between the AB system and C, with coherent velocities linking the proposed donor and recipient. CO observations could locate and quantify the molecular reservoir associated with the recent activity in C. These measurements would directly test the proposed connection between tidal gas removal from the satellites and renewed star formation in the primary massive galaxy. High-resolution observations with \textit{JWST} \citep{Gardner2006} and \textit{ALMA} \citep{Wootten2009} would be valuable for mapping the spatial distribution, phase structure, and kinematics of the gas reservoir associated with this weak rejuvenation event.

\subsection{The MW-LMC-SMC analogue}\label{Section5.3}
J1611+4026 provides an external analogue of the MW-- LMC-- SMC configuration: a massive primary galaxy accompanied by two interacting satellites. In this comparison, component C represents the primary, while components A and B constitute the satellite pair. Models of the Magellanic system illustrate how interactions between satellites can redistribute their gas before their eventual incorporation into the primary galaxy \citep{Besla2007, Besla2012}. Dynamical calculations also predict the eventual coalescence of the LMC with the MW \citep{Cautun2019}, placing the present satellite interaction within a longer sequence of hierarchical assembly.

J1611+4026 extends this comparison to a system with more massive companions relative to its host. Its combined satellite-to-primary stellar mass ratio, $(M_{\star,A}+M_{\star,B})/M_{\star,C}\simeq0.4$, exceeds that of the Magellanic system \citep{Besla2012, Bland-Hawthorn2016}. Moreover, component C is an ETG dominated by an old stellar population, whereas the MW is an actively star-forming disk galaxy. This difference makes J1611+4026 particularly useful for examining how interactions among satellites may influence a host whose recent star formation has otherwise been strongly suppressed.

The external perspective allows the satellite pair, the tidal structure associated with A, and the stellar population of C to be examined within a common projected spatial framework. Our internal view of the MW provides detailed information about individual stars, but recovering its integrated properties requires accounting for extinction, distances, and survey selection functions \citep{Rix2013, Bland-Hawthorn2016}. Recent reconstructions of its integrated stellar metallicity profile and radial light distribution demonstrate how these observations can be translated into quantities comparable to those measured in external galaxies \citep{Lian2023NatAs, Lian2024NatAs}. For J1611+4026, morphological decomposition and stellar population modeling instead allow the projected light distributions and integrated properties of the three components to be analyzed together. This connects the observed location of tidal disturbance in the satellite pair with the inferred recent stellar activity in the host.

The small burst mass fraction inferred for component C, $f_{\mathrm{burst}}<0.1\%$, quantifies the limited stellar mass contribution of its recent episode at the observed interaction stage. The system therefore combines clear evidence of satellite--satellite interaction with a young stellar component that contributes little to the mass of the primary. Under the gas-transfer interpretation discussed in \hyperref[Section5.2.3]{Section 5.2.3}, this configuration represents a stage in which material redistributed by the satellite encounter fuels primary massive galaxy rejuvenation. Consequently, J1611+4026 provides an observational benchmark for models linking the evolution of interacting satellite pairs to renewed star formation in a previously quiescent massive galaxy.

\section{Summary}\label{Section6}
We present a multi-wavelength study of J1611+4026, a unique triple galaxy system at $z\simeq0.14$, combining DESI and P200 spectroscopy with DESI-LS imaging and archival photometry. Image decomposition with \textsc{GALIGHT}, SED modeling with \textsc{CIGALE}, and stellar population fitting with \textsc{BAGPIPES} characterize the three components and reveal recent of star formation in the massive early-type host. Our main findings are as follows:

\begin{enumerate}
    \item Morphological decomposition reveals an asymmetric tidal feature extending from component A, providing clear evidence for an interaction between components A and B. Adopting the $ g$-, $ r$-, and $z$-band stellar mass-to-light ratios of component A, residual photometry yields a stellar mass of approximately $1.41\times10^{9}\,M_{\odot}$ for the detected portion of the tail. Spectroscopy identifies both satellites as SFGs containing chemically enriched gas.

    \item Component C is dominated by an old stellar population but exhibits substantial dust attenuation and an NUV excess. The \textsc{CIGALE} fit yields a nebular color-excess parameter of $E(B-V)_{\mathrm{lines}}=0.53\pm0.15$. The SFH reconstructed from the DESI optical spectrum with \textsc{BAGPIPES} indicates renewed star formation within the past $\sim100\,\mathrm{Myr}$ following an extended period of suppressed activity. The inferred recent episode contributes $f_{\mathrm{burst}}<0.1\%$ of the total stellar mass, identifying it as minor rejuvenation: a small addition of young stars to a predominantly old galaxy. 

    \item We propose that gas stripped during the interaction between A and B subsequently fuels the recent star formation in C. The tidal disturbance, the proximity of the satellite pair to the host, and the enriched gas present in both satellites identify the AB system as a plausible external source of fuel. In this interpretation, gas redistribution supplies material for host rejuvenation while the three galaxies remain distinct, linking satellite--satellite interaction to the recent evolution of the primary.

    \item The configuration of J1611+4026 provides an external analogue of the MW--LMC--SMC arrangement, extending this comparison to an early-type host with a larger relative stellar mass in its companions, $(M_{\star,A}+M_{\star,B})/M_{\star,C}\simeq0.4$. Its external perspective allows the satellite interaction, tidal structure, and host stellar population to be examined within a common projected spatial framework. The coexistence of a disturbed satellite pair and a small recent stellar mass contribution in C provides a quantitative benchmark for models connecting satellite interactions to host rejuvenation.
\end{enumerate}

In summary, the J1611+4026 system provides a valuable case study of how satellite interactions in group environments may drive the minor rejuvenation of massive galaxies through gas stripping and subsequent accretion. In the future, spatially resolved H~I and CO observations would directly test the proposed gas-supply pathway by tracing material between the satellite pair and C and measuring the molecular reservoir associated with the host's recent activity.

\section*{Data availability}
The optical images used in this article are available on the DESI Legacy Imaging Surveys website (\url{https://www.legacysurvey.org/}).

\begin{acknowledgements}
    We acknowledge support by the National Natural Science Foundation of China (No. 12473021), the National Key R\&D Program of China (No. 2024YFA1611602), and the Key Laboratory of Survey Science of Yunnan Province (No. 202449CE340002). Z.H. acknowledges support from the National Natural Science Foundation of China (Grant No. 12403104). This work is supported by the Telescope Access Program (TAP), which has been funded by the National Astronomical Observatories of China, the Chinese Academy of Sciences, and the Special Fund for Astronomy from the Ministry of Finance. The author thanks Prof. Xuheng Ding for helping with running \textsc{CIGALE}, and thanks to Yijun Wang for her help with better cartography. The author thanks Xingyu Zhu for the photometry redshift assistance provided earlier. The Legacy Surveys consist of three individual and complementary projects: the Dark Energy Camera Legacy Survey (DECaLS), the Beijing-Arizona Sky Survey (BASS), and the Mayall z-band Legacy Survey (MzLS). DECaLS, BASS, and MzLS together include data obtained, respectively, at the Blanco telescope, Cerro Tololo Inter-American Observatory, NSF’s NOIRLab; the Bok telescope, Steward Observatory, University of Arizona; and the Mayall telescope, Kitt Peak National Observatory, NOIRLab. The Legacy Surveys imaging of the DESI footprint is supported by the Director, Office of Science, Office of High Energy Physics of the U.S. Department of Energy under Contract No. DE-AC02-05CH1123, by the National Energy Research Scientific Computing Center, a DOE Office of Science User Facility under the same contract; and by the U.S. National Science Foundation, Division of Astronomical Sciences under Contract No. AST-0950945 to NOAO. Observations were obtained with the Hale Telescope at Palomar Observatory. This work made use of the code \textsc{CIGALE} (Code Investigating GALaxy Emission; \citealt{Boquien2019}. We acknowledge the use of the \textsc{BAGPIPES} code (Bayesian Analysis of Galaxies for Physical Inference and Parameter Estimation; \citealt{Carnall2018}).
\end{acknowledgements}

\bibliographystyle{aa}
\bibliography{refer}

\begin{appendix}

\onecolumn

\section{Spectral fitting by \textsc{QSOFITMORE}}
\label{AppendixA}

\begin{figure}[!htbp]
    \centering
    \includegraphics[
        width=0.9\linewidth,
        height=0.65\textheight,
        keepaspectratio
    ]{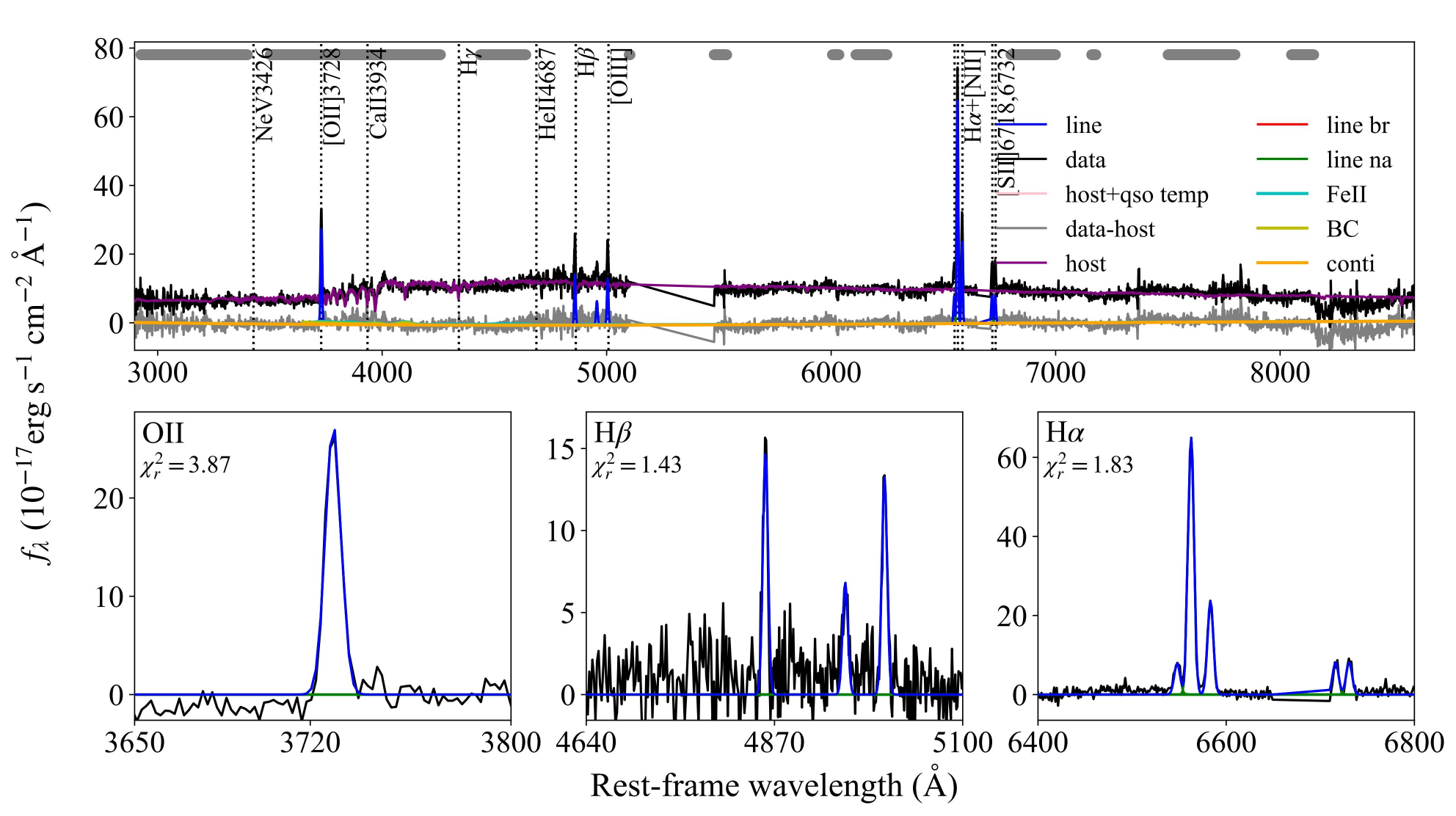}
    \caption{\textsc{QSOFITMORE} spectral fitting of J1611+4026A. Upper panel: The fitting of the full spectrum. Spectral regions covering 5105--5478{\AA} and 6650--6710{\AA} were masked during the fitting to exclude the bad CCD region of the DBSP red arm and telluric absorptions, respectively. Lower panels: Zoom in on the fitting of O II, $H{\alpha}$, and $H{\beta}$, marked with their corresponding $\chi^2_r$.}
    \label{appendix1}
\end{figure}

\begin{figure}[!htbp]
    \centering
    \includegraphics[
        width=0.9\linewidth,
        height=0.65\textheight,
        keepaspectratio
    ]{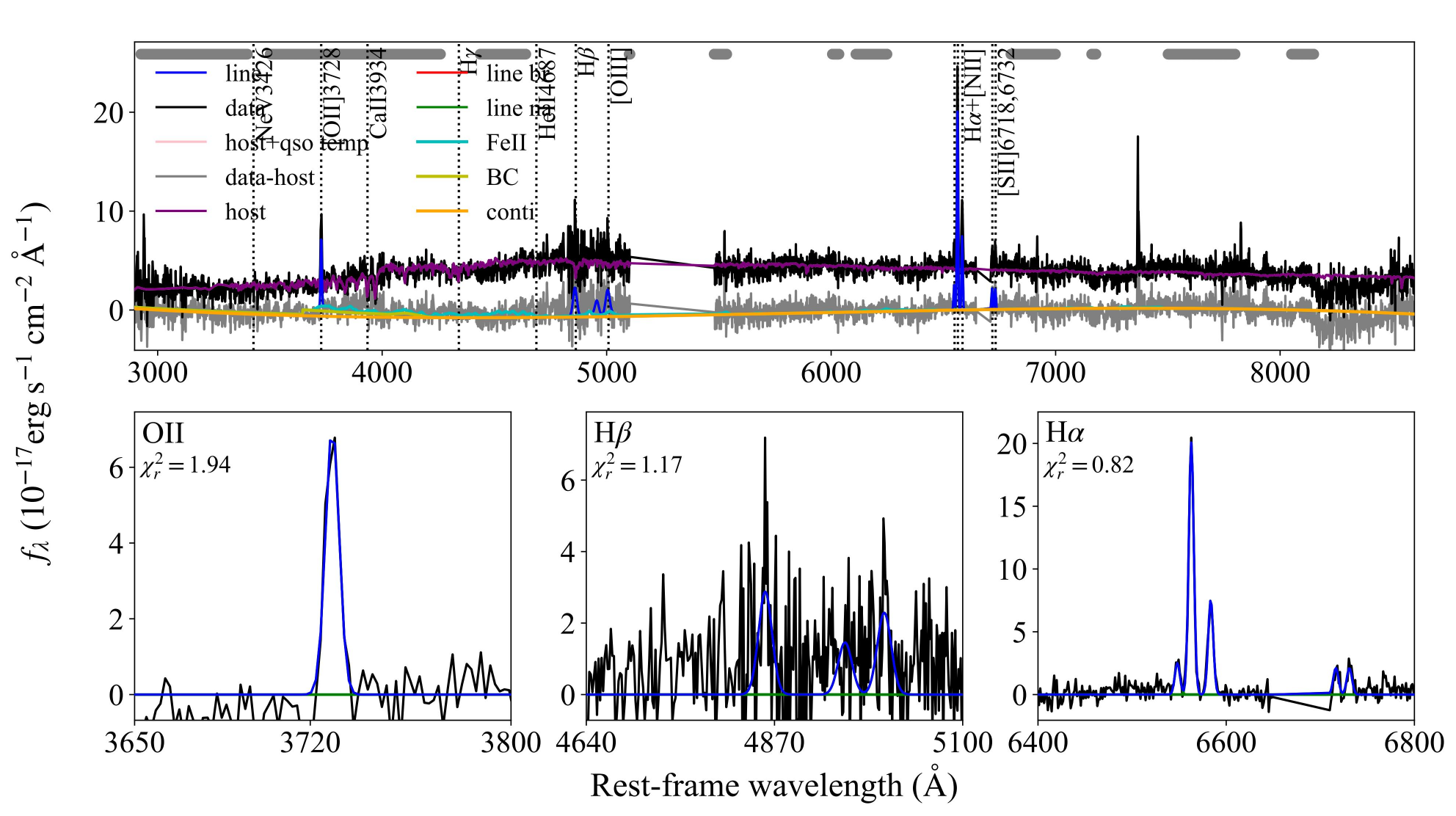}
    \caption{Same as \hyperref[appendix1]{Fig. A.1} but for J1611+4026B.}
    \label{appendix2}
\end{figure}

\begin{figure}[!htbp]
    \centering
    \includegraphics[
        width=0.9\linewidth,
        height=0.65\textheight,
        keepaspectratio
    ]{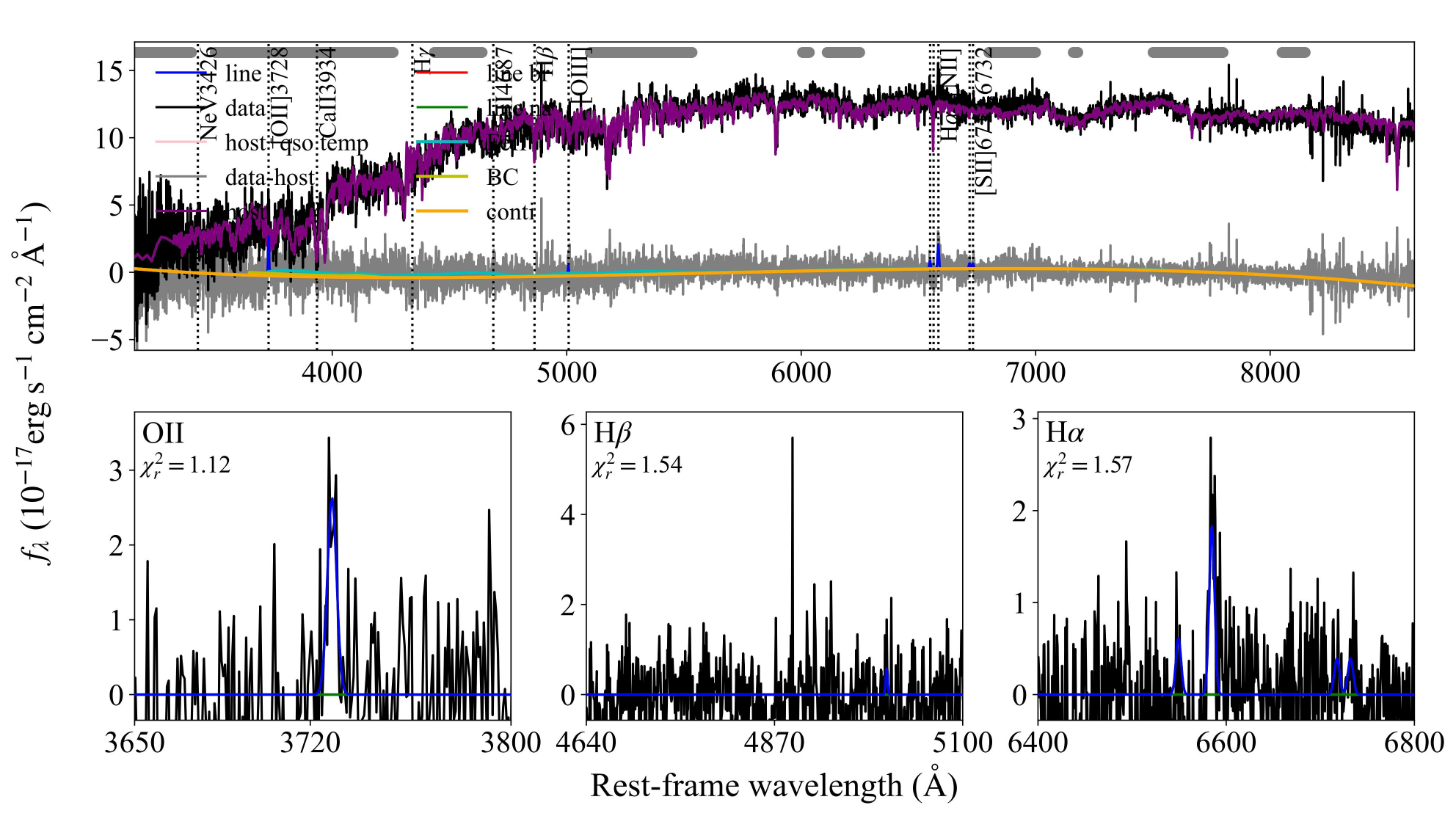}
    \caption{Similar to \hyperref[appendix1]{Fig. A.1} but for J1611+4026C, whose fitting results of SFR was unreliable.}
    \label{appendix3}
\end{figure}

The \textsc{QSOFITMORE} fitting results of components A, B, and C are shown in \hyperref[appendix1]{Fig. A.1}, \hyperref[appendix2]{Fig. A.2}, and \hyperref[appendix3]{Fig. A.3}, respectively. The fitting of component C was unreliable, as the selected templates might not be suitable for this component. We fit the broadband SED of component C using \textsc{CIGALE} to derive its SFR (see \hyperref[Section3.3]{Section 3.3}).

\clearpage

\section{Posterior analysis with \textsc{BAGPIPES}}
\label{AppendixB}

We present the joint posterior probability distribution derived from our independent \textsc{BAGPIPES} analysis to demonstrate the robustness of the derived dust attenuation and to explicitly rule out the age--dust degeneracy discussed in \hyperref[Section4.1]{Section 4.1}.

\begin{figure}[!htbp]
    \centering
    \includegraphics[
        width=0.70\linewidth,
        height=0.60\textheight,
        keepaspectratio
    ]{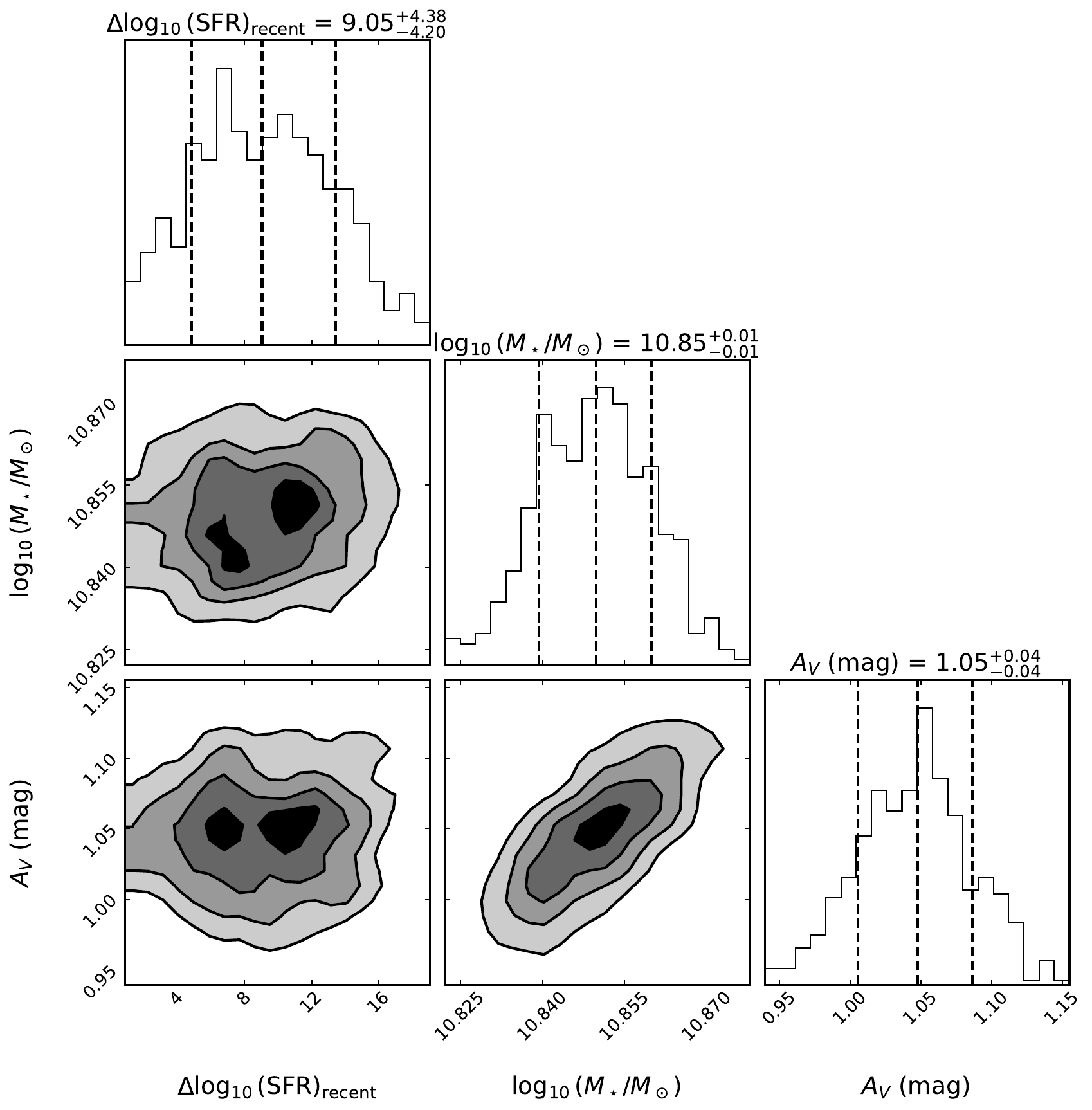}
    \caption{Posterior distributions of the recent SFR contrast, current stellar mass, and dust attenuation ($A_V$) inferred from the \textsc{BAGPIPES} fit to the DESI optical spectrum of component C. The quantity $\Delta\log_{10}(\mathrm{SFR})_{\mathrm{recent}}$ denotes the logarithmic ratio of the mean SFR over the past $100\,\mathrm{Myr}$ to that in the preceding $100$--$300\,\mathrm{Myr}$ interval. Diagonal panels show the marginalized distributions, with dashed lines marking the 16th, 50th, and 84th percentiles; off-diagonal panels show the joint posterior distributions. The posterior favors enhanced recent star formation. The recent SFR contrast exhibits weak monotonic correlations with $A_V$ and stellar mass, with Spearman coefficients of $\rho_{\mathrm{s}}=0.053$ and $0.097$, respectively.}
    \label{appendix4}
\end{figure}

\clearpage

\section{Input parameters of SED modelling}
\label{AppendixC}

We present the detailed input parameters and their corresponding ranges adopted for the SED fitting with \textsc{CIGALE}.

\begin{table}[!htbp]
    \centering
    \caption{Input parameters for the SED fitting by \textsc{CIGALE}.}
    \label{tab_SED}

    \small
    \renewcommand{\arraystretch}{1.2}

    \begin{tabularx}{\linewidth}{
        @{}
        >{\raggedright\arraybackslash}p{0.23\linewidth}
        >{\raggedright\arraybackslash}p{0.25\linewidth}
        >{\raggedright\arraybackslash}p{0.15\linewidth}
        >{\raggedright\arraybackslash}X
        @{}
    }
        \hline
        \hline
        Module & Parameter & Symbol & Values \\
        \hline

        \textbf{Star formation history}
        & e-folding time of main population
        & $\tau_{\text{main}}$ (Myr)
        & 1000, 2000, 3000, 5000, 8000 \\

        (Delayed SFH + Burst/Quench)
        & Age of main population
        & $t_{\text{main}}$ (Myr)
        & 2000, 5000, 8000, 10000 \\

        & Age of burst/quench episode
        & $t_{\text{burst}}$ (Myr)
        & 5, 10, 20, 50, 100, 300, 500, 1000 \\

        \hline

        \textbf{Simple stellar population}
        & Initial mass function
        & IMF
        & \citep{Salpeter1955} \\

        \citep{Bruzual2003}
        & Metallicity
        & $Z$
        & 0.02 ($Z_\odot$), 0.05 \\

        \hline

        \textbf{Dust attenuation}
        & Color excess of nebular lines
        & $E(B-V)_{\text{lines}}$ (mag)
        & 0.05, 0.1, 0.2, \ldots, 0.9 \\

        \citep{Calzetti2000}
        & Stellar continuum factor
        & $f_{\text{att}}$
        & 0.44 \\

        \hline

        \textbf{Dust emission}
        & Dust Temperature
        & $T_{\text{dust}}$ (K)
        & 20, 25, 30, 35, 40, 45, 50 \\

        \citep{Casey2012}
        & Emissivity index
        & $\beta$
        & 1.6 \\

        & MIR power law slope
        & $\alpha$
        & 2.0 \\

        \hline

        \textbf{AGN emission}
        & AGN fraction
        & $f_{\text{AGN}}$
        & 0.0, 0.02, 0.05, 0.1, \ldots, 0.99 \\

        \citep[SKIRTOR,][]{Stalevski2016}
        & Viewing angle
        & $i$ (deg)
        & 30, 70 \\

        & Optical depth (9.7 $\mu$m)
        & $\tau_{9.7}$
        & 3, 5, 7 \\

        \hline
    \end{tabularx}

    \tablefoot{
        $f_{\text{att}}
        = E(B-V)_{\star}/E(B-V)_{\text{lines}}$.
    }
\end{table}

\clearpage
\end{appendix}

\end{document}